\documentstyle[psfig]{article}
\makeatletter
\def\@aabuffer{}
\def\author #1{\expandafter\def\expandafter\@aabuffer\expandafter
{\@aabuffer \normalsize\rm      #1\relax \par\vspace{0.75em}}}
\def\address#1{\expandafter\def\expandafter\@aabuffer\expandafter
{\@aabuffer \normalsize\it #1\relax \par\vspace{1.5em}}}

\def\maketitle{
\begin{center}
   {\bf \@title \par}        
   \vskip 2em                      % Vertical space after title.
   \@aabuffer\relax
\end{center} \par
\gdef\@aabuffer{}
}

\def\abstracts#1{
\centering{\begin{minipage}{5.75truein}
                 \footnotesize
                 \parindent=0pt #1\par
                 \end{minipage}}
                 \vskip 2em \par}

\twocolumn
\def\section{\@startsection {section}{1}{\z@}{-3.5ex plus -1ex minus 
    -.2ex}{2.3ex plus .2ex}{\bf }}
\def\subsection{\@startsection{subsection}{2}{\z@}{-3.25ex plus -1ex minus 
   -.2ex}{1.5ex plus .2ex}{\it }}

\def\@makefnmark{{$\!^{\@thefnmark}$}}

\renewenvironment{thebibliography}[1]
	{\begin{list}{\arabic{enumi}.}
	{\usecounter{enumi}\setlength{\parsep}{0pt}
	 \setlength{\itemsep}{0pt} \settowidth
	{\labelwidth}{#1.}\sloppy}}{\end{list}}

\newcounter{arabiclistc}

\def\@citex[#1]#2{\if@filesw\immediate\write\@auxout
	{\string\citation{#2}}\fi
\def\@citea{}\@cite{\@for\@citeb:=#2\do
	{\@citea\def\@citea{,}\@ifundefined
	{b@\@citeb}{{\bf ?}\@warning
	{Citation `\@citeb' on page \thepage \space undefined}}
	{\csname b@\@citeb\endcsname}}}{#1}}

\newif\if@cghi
\def\cite{\@cghitrue\@ifnextchar [{\@tempswatrue
	\@citex}{\@tempswafalse\@citex[]}}
\def\citelow{\@cghifalse\@ifnextchar [{\@tempswatrue
	\@citex}{\@tempswafalse\@citex[]}}
\def\@cite#1#2{{$\!^{#1}$\if@tempswa\typeout
	{IJCGA warning: optional citation argument 
	ignored: `#2'} \fi}}

\def\baselinestretch{1.0}
\ifx\selectfont\undefined
\@normalsize\else\let\glb@currsize=\relax\selectfont
\fi

\ifx\selectfont\undefined
\def\@singlespacing{%
\def\baselinestretch{1}\ifx\@currsize\normalsize\@normalsize\else\@currsize\fi%
}
\else
\def\@singlespacing{\def\baselinestretch{1}\let\glb@currsize=\relax\selectfont}
\fi

\long\def\@makecaption#1#2{
   \vskip 10pt 
   \setbox\@tempboxa\hbox{\footnotesize #1: #2}
   \ifdim \wd\@tempboxa >\hsize   % IF longer than one line:
       \leftskip 0pt plus 1fil 
       \rightskip 0pt plus -1fil 
       \parfillskip 0pt plus 2fil 
       \footnotesize #1: #2\par   %   THEN set as ordinary paragraph.
     \else                        %   ELSE  center.
       \hbox to\hsize{\hfil\box\@tempboxa\hfil}  
   \fi}

\makeatother

\def\Journal#1#2#3#4{{#1} {\bf #2}, #3 (#4)}

\def\NPB{{\em Nucl. Phys.} B}
\def\PLB{{\em Phys. Lett.}  B}
\def\PRL{\em Phys. Rev. Lett.}
\def\PRD{{\em Phys. Rev.} D}
\def\ZPC{{\em Z. Phys.} C}

\def\mco{\multicolumn}

\def\be{\begin{equation}}
\def\ee{\end{equation}}
\def\bea{\begin{eqnarray}}
\def\eea{\end{eqnarray}}
\begin{document}
\title{~\hfill LBNL-39484\\
~\hfill UCB-PTH-96/44\\
NUCLEON DECAY IN GUT AND NON-GUT SUSY MODELS}

\author{ HITOSHI MURAYAMA}

\address{Department of Physics, University of California\\
Berkeley, CA 94720, USA}
\address{Theoretical Physics Group, Lawrence Berkeley National Laboratory\\
University of California, Berkeley, CA 94720, USA}

%%%%%%%%%%%%%%%%%%%%%%%%%%%%%%%%%%%%%%%%%%%%%%%%%%%%%%%%%%%%%%
% You may repeat \author \address as often as necessary      %
%%%%%%%%%%%%%%%%%%%%%%%%%%%%%%%%%%%%%%%%%%%%%%%%%%%%%%%%%%%%%%

\twocolumn[\maketitle\abstracts{
I first emphasize the importance of searching for nucleon decay in the
context of supersymmetric models.  The status of minimal SUSY SU(5)
model is reviewed, which can be definitively ruled out by a combination
of superKamiokande and LEP-2 experiments.  Non-minimal models may
provide some suppression in the nucleon decay rates, but there is still
a good chance for superKamiokande.  I point out that the operators
suppressed even by the Planck-scale are too large.  We need a
suppression mechanism for the operators at the level of $10^{-7}$, and
the mechanism, I argue, may well be a flavor symmetry.  A particular
example predicts $p \rightarrow K^0 e^+$ to be the dominant mode which
does not arise in GUT models.
}]

\section{Introduction}

Now superKamiokande is up and running very well!  This is the good news
which we heard at this conference.  And it is expected to extend the
reach on nucleon decay by more than an order of magnitude.  My talk is
devoted to discuss the following questions about the nucleon decay in
the context of suprsymmetric models.  How
important is it to look for nucleon decay?  What decay modes are
expected or interesting?  What is the current status of various models
which predict nucleon decay?

\section{Why Nucleon Decay?}

Here I would like to remind you why it is so important and exciting to 
look for nucleon decay experimentally.  

There are at least three reasons why nucleons may decay.  First, we 
have seen a dramatic success of supersymmetric grand unified theory 
(SUSY-GUT) in predicting $\sin^{2}\theta_{W}$.  If we take this hint 
seriously, we expect to see nucleon decay since SUSY-GUTs predict 
nucleon decay at an observable rate.  Second, the quantum gravity 
effects are believed to break any global symmetries, and hence the 
baryon and/or lepton numbers are also likely to be broken.  Third, we 
know that our Universe is dominantly made up of baryons rather than 
anti-baryons with a possible exception inside the Tevatron ring.  If 
we would like to understand this asymmetry as a result of dynamics in 
the Early Universe, there must be interactions which violate baryon 
number conservation which were effective at high temperatures.  

I must admit none of the above arguments are without loopholes.  First 
motivation based on the apparent gauge unification may not necessarily 
mean that there is a field theoretical grand unification.  It may be 
explained by, for instance, string unification where you do not have a 
simple large gauge group into which the standard model gauge groups 
are embedded.  The second argument may not necessarily imply the 
existence of nucleon decay process.  The baryon number may be
effectively preserved due to gauge symmetries, either continuous 
(such as gauged U(1)$_{B}$ symmetry\,\cite{U(1)B}) or
discrete,\cite{discreteB} which are stable against quantum  
gravitational effects.  Finally, the last motivation may be void if the 
cosmic baryon asymmetry is generated due to the sphaleron effect from 
the primordial lepton asymmetry, which may be generated due to the 
decay of right-handed neutrinos\,\cite{FY} or Affleck--Dine mechanism.\cite{MY}

However, I should also emphasize the following simple fact which by 
itself makes the search for nucleon decay very interesting in the 
supersymmetric models.  We are probing physics at extremely high energy 
scales by looking for nucleon decay.  In particular, the current limit 
on nucleon decay has a sensitivity up to $10^{26}$~GeV!  Of course it 
does not make much sense to talk about such a scale much beyond the 
Planck scale.  I will explain below where this scale comes from.  In 
any case, such an extreme sensitivity to high-energy physics is hard 
to beat by any other means, and this is what makes the nucleon decay 
such an interesting process to look for.

\section{$D=5$ or $D=6$}

Let me briefly discuss the ``classic'' prediction of a grand unified 
theory how a nucleon might decay.  In SU(5) GUT, the standard model 
gauge groups are embedded into a simple SU(5) group which has 
additional gauge bosons beyond those in the standard model.  The 
additional gauge bosons mediate a process such as $uu\rightarrow 
e^{+}\bar{d}$.  This process can be effectively described by a $D=6$ 
four-fermion operator
\begin{equation}
{\cal L} = \frac{1}{M^2} uude ,
\end{equation}
where $M$ is a high mass scale such as the GUT-scale. 
By adding another down quark as a spectator to this 
process, one obtains a decay $p\rightarrow e^{+} \pi^{0}$.  The 
current lower limit on proton partial lifetime implies the GUT-scale 
must be larger than $1.5\times 10^{15}$~GeV where I estimated the 
bound conservatively using formulae given in.\cite{HMY}  Because the 
operator has a suppression by two powers of a high mass scale, the 
proton decay rate is given roughly by $\Gamma_{p} \sim 
m_{p}^{5}/M^{4}$ and suppressed by the fourth power.  It is not easy 
to extend the reach to higher mass scale in this case.

On the other hand, supersymmetric models tend to have operators which 
mediate nucleon decay with less suppression by a high mass scale, such 
as
\begin{equation}
{\cal L} = \frac{\lambda}{M} q q \tilde{q} \tilde{l}
\end{equation}
with $\lambda$ a coupling constant.  This type of operators has $D=5$
and they are called $D=5$ operators.  The squark/slepton created
virtually by this type of operators must be converted to quark/lepton by
an exchange of gauginos to let a nucleon decay.  As a result, the
nucleon decay rate is given roughly by $\Gamma_{p} \sim \lambda^{2}
m_{p}^{5}/M^{2}/m_{SUSY}^{2}$.  Since it is suppressed only by two
powers in a high mass scale, we can probe much higher $M$ with these
operators.  In fact, if we take $\lambda \simeq 1$, and by doing an
extremely conservative analysis as the one which I will describe shortly
in the case of minimal SUSY SU(5) GUT, one obtains a lower bound on $M$:
\begin{equation}
	M > 8\times 10^{23}~\mbox{GeV}.
\end{equation}

Possible existence of such a $D=5$ operator was first pointed out in the
context of SUSY-GUT.\cite{D=5}  When the standard model gauge groups are
embedded into SU(5), the Higgs doublets $H$ which break the electroweak
symmetry are embedded into {\bf 5} and {\bf 5}$^{*}$ representations of
SU(5) which contain color-triplet Higgs bosons $H_{C}$.  They further
have their fermionic partners $\tilde{H}_{C}$ due to supersymmetry.  The
exchange of color-triplet Higgsinos generate operators suppressed only
by one power in $M_{GUT}$ because of the fermion propagator $\sim
i/(\not\!p - M)$.  Since the couplings of color-triplet Higgsinos to
(s)quarks and (s)leptons are related to those of color-triplet Higgs
bosons by supersymmetry and further to those of Higgs doublets by SU(5),
we know the strengths of the couplings rather well.  The most important
$D=5$ operator has a coefficient $\lambda_{c}\lambda_{s}
\sin \theta_{C}/M_{H_C}$ where $\lambda_{c}$, $\lambda_{s}$ are the 
Yukawa couplings of charm and strange quarks to the standard Higgs 
bosons and $\theta_{C}$ the Cabbibo angle.  Therefore we can make 
precise predictions of nucleon decay rate in this situation for given 
values of $M_{H_{C}}$.  

On the other hand, the quantum gravity effect may well generate 
effective non-renormalizable operators if they break global baryon 
and/or lepton number symmetry.  They are likely to be suppressed by 
powers in the reduced Planck scale $M_{*} \equiv M_{Pl}/\sqrt{8\pi}$ 
because they are quantum gravity effects.  They may arise also due to 
the exchange of heavy string states.  Unless there is a reason for a 
suppression, we expect the coefficient of a $D=5$ operator to be 
$1/M_{*}$ in this case.

I will discuss the consequence of $D=5$ operators of GUT- and Planck-scale 
origin separately in the following sections.

Whatever the origin of a $D=5$ operator is, there are a couple of 
characteristics common to nucleon decay via $D=5$ operators.  (1) It 
is sensitive to {\it extremely}\/ high energy scales, as already 
mentioned above.  Actually it is a phenomenological disaster if there 
is an operator with a coefficient of order $1/M_{*}$. Therefore, the 
current bound is already putting constraints on the physics at the 
Planck scale.  (2)  The final states of nucleon decays (almost) 
always involve kaons, either $K^{+}$ or $K^{0}$.  This is due to the 
flavor SU(3) symmetry property of $D=5$ operators.  There is no $D=5$ 
operator which consists of the first generation fields only; it 
identically vanishes.  The quark which is not in the first generation 
but still light enough to be able to appear in the nucleon decay is 
the strange quark.  (3) The rate depends on the superparticle 
spectrum, such as masses of squarks, sleptons and wino.  

\section{Minimal SUSY SU(5)}

\begin{table}
\begin{center}
\caption{Relative decay rates of nucleons in the Minimal SUSY 
SU(5)-GUT assuming there is no accidental cancellation in the 
amplitudes.\label{tab:rel}}
\vspace{0.4cm}
\begin{tabular}{|c|c|c|c|c|}
\hline &&&&\\
$p \rightarrow$ & $K^{+} \bar{\nu}_{\mu}$ & $\pi^{+} \bar{\nu}_{\mu}$ & 
$K^{0} \mu^{+}$ & $K^{0} e^{+}$\\ 
rel. rates & 1 & 0.49 & 0.00069 & $2.1\cdot 10^{-6}$\\ \hline
$n \rightarrow$ & \mco{2}{|c|}{$K^{0} \bar{\nu}_{\mu}$} &
\mco{2}{|c|}{$\pi^{0} \bar{\nu}_{\mu}$} \\
rel. rates & \mco{2}{|c|}{1.8} & \mco{2}{|c|}{0.24}
\\ \hline
\end{tabular}
\end{center}
\end{table}

The nucleon decay rate can be worked out quantitatively\,\cite{HMY} in
the Minimal SUSY SU(5)-GUT.\cite{minimal} As explained already, the
$D=5$ operators arise because of the exchange of the color-triplet
Higgs(ino), and the dominant operator has the coefficient $\lambda_{c}
\lambda_{s} \sin\theta_{C}/ M_{H_{C}}$.  There are 
four types of parameters which enter the calculation.  (1) The Yukawa 
couplings $\lambda_{c}$ and $\lambda_{s}$ are known up to the 
dependence on $\tan\beta$.  The amplitude is proportional to 
$1/\sin2\beta$ which grows with $\tan\beta$.  (2) The mass of the 
color-triplet Higgs $M_{H_{C}}$ can be actually determined from the 
low-energy data only, namely the gauge coupling constants measured by 
LEP. At the one-loop level, it can be determined by the following 
formula\,\cite{HMY1}
\begin{eqnarray}
\lefteqn{
	(3 \alpha^{-1}_{2} - 2\alpha^{-1}_{3} - \alpha^{-1}_{1})(m_{Z})}
		& & \nonumber \\
& &
	= \frac{1}{2\pi} \left( \frac{12}{5} \ln \frac{M_{H_{C}}}{m_{Z}}
		- 2 \ln \frac{m_{SUSY}}{m_{Z}} \right) .
\label{eq:HMY1}
\end{eqnarray}
The largest uncertainty is in $\alpha_{3}(m_{Z})$, and 
therefore we put bounds as a function of $\alpha_{3}(m_{Z})$.  In practice, I 
use two-loop renormalization group equations.  Note the positive 
correlation between $\alpha_{3} (m_{Z})$ and $M_{H_{C}}$: the decay rate is 
larger for smaller $\alpha_{3}(m_{Z})$.  (3) We choose the most 
conservative choice of the superparticle mass spectrum which gives the 
smallest nucleon decay rate.  The amplitude is proportional to 
$M_{2}/m_{\tilde{q}}^{2}$, and we take $M_{2} \simeq 45$~GeV and 
$m_{\tilde{q}} \simeq 1$~TeV. (4) The matrix element $\beta$ of the 
operator between a nucleon and a meson is not well known.  The 
estimates vary as $\beta = 0.003$--0.03~GeV$^{3}$.\cite{BEHS} We 
again take the most conservative one $\beta=0.003$~GeV$^{3}$.  (5) We 
use the subdominant decay mode $n \rightarrow \pi^{0} \bar{\nu}_{\mu}$ 
instead of the dominant one $n \rightarrow K^{0} \bar{\nu}_{\mu}$ 
because there might be a partial cancellation in the amplitude between 
a diagram with charm (s)quark and one with top (s)quark.\cite{NA}  For
the case  without such a cancellation, the relative decay ratios are
given in the Table~\ref{tab:rel}.  

\begin{figure}
\centerline{
\psfig{figure=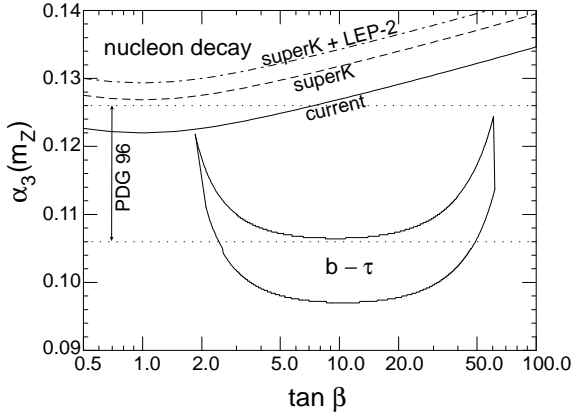,width=3in}
}
\caption{Excluded region on $(\tan\beta, \alpha_3(m_Z))$ space from
nucleon decay based on very conservative assumptions as described in the
text.  Expected improvements from superKamiokande and LEP-2 are also
shown.  The range shown for $\alpha_3 (m_Z)$ from PDG96 is two sigma
range.  The preferred region from $b-\tau$ unification is also shown for
$m_t = 176$~GeV.}
\label{fig:CM}
\end{figure}

Based on very conservative assumptions as describe above, the allowed 
region\,\cite{CM} in $(\tan\beta, \alpha_{3} (m_{Z}))$ plane is given in 
Fig.~\ref{fig:CM}.  The experimental limit on nucleon decay puts a 
lower bound on $M_{H_{C}}$ which is translated into the lower limit 
on $\alpha_{3}(m_{Z})$ using the correlation.  The bound is tighter 
for larger $\tan\beta$ because the amplitude grows.  The two-sigma 
band of $\alpha_{3} (m_{Z})= 0.118 \pm 0.003$ is shown.  For a 
comparison, the preferred range from $b$-$\tau$ Yukawa unification is 
also shown for $m_{t} = 176$~GeV.  They actually barely overlap for smaller 
$m_{t}$.  The expected improvements by superKamiokande (dashed) and 
further by LEP-2 (dotted)  are also shown.  (An improved lower limit 
on $M_{2}$ from LEP-2 would make the amplitude larger $\sim 
M_{2}/m_{\tilde{q}}^{2}$.)  The Minimal SUSY SU(5)-GUT can be 
definitively excluded by these experiments.  

\section{Non-minimal SUSY-GUT}

There are many good reasons to discuss extensions of the minimal SUSY
SU(5) GUT.  Among them, there are two points directly relevant to the
nucleon decay.  (1) The triplet-doublet splitting problem.  In minimal
SUSY SU(5) GUT, one needs to fine-tune independent parameters at the
level of $10^{-14}$ to keep Higgs doublets light while making the
color-triplet Higgs heavy.  (2) The wrong fermion mass relations.  It
predicts $m_s = m_\mu$ and $m_d = m_e$ at the GUT-scale, which are off
from the phenomenologically preferred Georgi--Jarlskog relations $m_s =
m_\mu/3$, $m_d = 3 m_e$.  

Solutions to the above-mentioned problems modify the predicted rate and
branching ratios of the nucleon decay.  One possible attempt to obtain
Georgi--Jarlskog relations is to use the SU(5)-adjoint Higgs to
construct an effective {\bf 45} Higgs doublets as composites of ordinary
Higgs doublets in {\bf 5} and the adjoint.  This modification leads to a
factor-of-two enhancement in the amplitude; a factor of four in the
rate.\cite{INS}  The relative branching ratios can be also different.
It remains true that the $K^{+,0} \bar{\nu}_\mu$ modes are the dominant
ones, while the $K^0 \mu^+$ mode may be much less suppressed than in the
minimal SU(5).\cite{Antaramian,BB}

There are various proposals to solve the triplet-doublet splitting 
problem, which lead to completely different nucleon decay 
phenomenology.  I discuss three of them here.  (1) The missing partner 
model,\cite{MNTY} (2) Dimopoulos--Wilczek--Srednicki 
mechanism,\cite{DWS} and (3) flipped SU(5) model.\cite{flipped}

In the missing partner model, one employs {\bf 75} representation to
break SU(5) instead of the adjoint {\bf 24}, and further introduces {\bf
50} and {\bf 50}$^*$ representations which mix with the
color-triplet Higgs to make them massive.  Since the model involves such
large representations, the size of the GUT-scale threshold corrections
are significantly larger than that in the minimal model.  And the
correction changes the determination of the color-triplet Higgs mass as
done in Eq.~(\ref{eq:HMY1}), and the measured values of the gauge
coupling constants prefer larger $M_{H_C}$ than in the minimal
model.\cite{Yamada}  In this case the proton decay rates are much more
suppressed, by a few orders of magnitudes.  One drawback of the model is
that it becomes non-perturbative well below the Planck scale due to
large representations and one needs to complicate the model further to
keep it perturbative.\cite{missingPQ}  It is worth to recall that the
minimal SU(5) model is marginally allowed only with very conservative
assumptions made in the previous section.   Even though there is an
additional suppression to the proton decay rate in this class of models,
the decay rate may still well be within the reach of superKamiokande
experiment.  

The mechanism proposed by Dimopoulos, Wilczek and further by Srednicki 
employs SO(10) unification with Higgs fields in adjoint and symmetric 
tensor representations which naturally keep Higgs doublets light.  
However, their model breaks SO(10) only to SU(3)$\times$SU(2)$_L 
\times$SU(2)$_R \times$U(1)$_{B-L}$ and has to be extended to achieve 
the desired symmetry breaking down to the standard model gauge group.  
One of such extensions by Babu and Barr\,\cite{BB0} eliminates $D=5$ 
operators entirely; but it involves rather complicated Higgs sector, and one 
needs to forbid some allowed interactions in the superpotential 
arbitrarily.  A later attempt\cite{BB1} to guarantee the special form of the 
superpotential by symmetries did not eliminate the $D=5$ operators 
entirely, but resulted in a weak suppression of the operators.  Again 
in view of the very marginal situation in the minimal model, the decay 
rate could be within the reach of the superKamiokande.

The flipped SU(5) model solves the triplet-doublet splitting problem in
a way that it also eliminates the $D=5$ operators entirely.  A possible
problem with this model is that the gauge unification becomes more or
less an accident rather than a prediction.  On the other hand, the
elimination of the $D=5$ operator is a natural consequence of the
structure of the Higgs sector, and is rather a robust prediction of the
model except the Planck-scale effects which will be discussed below.  An
interesting feature of the model is that the GUT-scale is determined by
$\alpha_2$ and $\alpha_3$ and hence can be {\it lower}\/ than the scale
in the minimal SU(5) which is determined by $\alpha_2$ and $\alpha_1$.
Since the model does not predict the relation between $\alpha_1$ and
$\alpha_{SU(5)}$, $\alpha_1$ does not need to meet with the other
coupling constants at the same scale.  Therefore, the GUT-scale can be
as low as $M_{GUT}^{\rm flipped} = 4$--$20\times 10^{15}$~GeV.  If the
$M_{GUT}$ is at the low side within this range, the $D=6$ operator may
be observable in the $\pi^0 e^+$ mode,\cite{ELN} since the superKamiokande
is expected to extend the reach by a factor of 20.

\section{Planck-scale Operators}

As I mentioned at the beginning of the talk, the Planck-scale effects
may generate $D=5$ operators suppressed by the reduced Planck scale $M_*
= 2\times 10^{18}$~GeV.  Even when there is no color-triplet Higgs, such
as in string compactifications which breaks the gauge group down to the
standard model (with possible U(1) factors) directly, the higher string
excitations may give rise to effective non-renormalizable $D=5$
operators which break baryon- and/or lepton-number symmetries.  For 
$D=5$ operators which involve first- and second-generation fields,
$1/M_*$ suppression is far from enough: one needs a coupling constant of
order $10^{-7}$ to keep the nucleons stable enough as required by
experiments.  

It is a serious question in supersymmetry phenomenology why the
Planck-scale $D=5$ operators are so much suppressed.  Even though there
are ways to forbid them by employing discrete gauge
symmetries,\cite{discreteB} I prefer a different type of solution:  the
$D=5$ operators are suppressed because of the same reason why the Yukawa
couplings of light generations are suppressed.\cite{MK}  One way to
understand why the Yukawa couplings are so small, such as $10^{-6}$ for
the case of the electron, may be a natural consequence of an approximate
flavor symmetry.  If a flavor symmetry exists and is only weakly broken
to explain smallness of the Yukawa couplings, the same flavor symmetry
can well suppress the $D=5$ operators at the Planck-scale.  We\,\cite{MK}
speculated that the $D=5$ operators with such a flavor origin may have
very different flavor structure from those in the GUT models, and may lead to
quite different decay modes like $p\rightarrow K^0 e^+$.

Hall and myself constructed a model with $(S_3)^3$
symmetry\,\cite{S3} in which the hierarchical Yukawa matrices can be
understood as a consequence of sequential breaking of the flavor
symmetry while the symmetry preserves sufficient degeneracy among the
squarks and sleptons to suppress flavor-changing neutral currents.  It
happened that the flavor symmetry in this model also suppresses $D=5$
operators to the level of about 1/9 of the minimal SU(5)model, so that
it can well be within the reach of superKamiokande.\cite{S3q}  What is
particularly interesting in this model is that it predicts $p
\rightarrow K^0 e^+$ as the {\it dominant}\/ mode over the $K^+
\bar{\nu}$, while $n\rightarrow K^0 \bar{\nu}_e$ is the dominant mode in
neutron decay with a comparable rate.  

Finally, I would like to make a brief comment on the $R$-parity.  The
$R$-parity is usually imposed on supersymmetric models as means to
forbid dangerous $D=4$ (!) operators which break baryon- or
lepton-numbers.  Indeed, such operators must be highly suppressed $<
10^{-13}$ if both $B$- and $L$-violating operators are present.
However, a flavor symmetry may suppress these operators drastically as
well,\cite{MK} and explicit examples were constructed.\cite{Rparity}

\section{Conclusion}

I reminded you the importance of search for nucleon decay in the context
of supersymmetric theories.  Contrary to the non-supersymmetric models,
nucleon decay probes physics even at the Planck-scale because of
possible $D=5$ operators.  In fact, the current limit requires the scale
of baryon number violation to be larger than $10^{24}$~GeV or more if
the operators are unsuppressed.  We expect $K\bar{\nu}$
modes to be dominant in many supersymmetric models.

The minimal SUSY SU(5)-GUT is nearly excluded by a combination of
$\alpha_3(m_Z)$ prediction and the nucleon decay, and will be
definitively with superKamiokande and LEP-2.  Non-minimal GUT models may
suppress the nucleon decay rates, but still the rates could be within
the reach of superKamiokande.  A dark horse example is the flipped
SU(5) model which may give $\pi^0 e^+$ mode at an observable rate.

Since the Planck-scale operators give nucleon decay rates which are too
large, one needs a suppression mechanism.  I argued that a flavor symmetry
is a likely mechanism to provide an adequate suppression of the $D=5$
operators, and they may give rise to exotic decay modes like $K^0 e^+$
which do not arise in GUT models.

\section*{Acknowledgments}
This work was supported in part by the Director, Office of 
Energy Research, Office of High Energy and Nuclear Physics, Division of 
High Energy Physics of the U.S. Department of Energy under Contract 
DE-AC03-76SF00098, by the National Science Foundation under 
grant PHY-95-14797, and also by the Alfred P. Sloan 
Foundation.

\end{document}